\documentclass[twocolumn,showpacs,showkeys]{revtex4}
\usepackage{graphicx,subfigure,epsfig}
\def\beq{\begin{equation}}
\def\eeq{\end{equation}}
\def\bea{\begin{eqnarray}}
\def\eea{\end{eqnarray}}
\def\d{{\mathrm{d}}}

\newfont{\cursive}{pzcmi at 9pt}
\def\~t{\tilde{t}}
\def\Painleve{Painlev\'{e}}

\def\e2phi{\e^{2\Phi}}

\begin{document}

\title{Spherically symmetric trapping horizons, the Misner-Sharp mass and black hole evaporation}

\author{Alex B. Nielsen $^1$  and Dong-han Yeom $^2$ \\} 
\affiliation{$^1$ Center for Theoretical Physics
and School of Physics, College of Natural Sciences
Seoul National University, Seoul 151-742, Korea }\email{eujin@phya.snu.ac.kr}

\affiliation{$^2$ Department of Physics, Korea Advanced Institute of Science and Technology, Daejeon 305-701, Korea}\email{innocent@muon.kaist.ac.kr}

\begin{abstract}
Understood in terms of pure states evolving into mixed states, the possibility of information loss in black holes is closely related to the global causal structure of spacetime, as is the existence of event horizons. However, black holes need not be defined by event horizons, and in fact we argue that in order to have a fully unitary evolution for black holes, they should be defined in terms of something else, such as a trapping horizon. The Misner-Sharp mass in spherical symmetry shows very simply how trapping horizons can give rise to black hole thermodynamics, Hawking radiation and singularities. We show how the Misner-Sharp mass can also be used to give insights into the process of collapse and evaporation of locally defined black holes.

\end{abstract}
\maketitle

\section{Introduction}

The possibility of information loss in black hole evaporation has been with us for many years. There have been many attempts over the years to resolve the issue one way or another, but it is probably fair to say that there is still no consensus on what the correct picture of a black hole and its evolution should be.

Indeed, the possibility that the evolution of a black hole may be a non-unitary process has many disturbing consequences. We will not attempt here to provide a definitive answer to the question of whether black hole evaporation is unitary or not. We believe that this is a question that can only be answered in the context of a theory that we do not yet possess. Instead what we want to do here is to highlight some of the semi-classical issues involved in the problem and the implicit assumptions that are often under-appreciated in the analysis.

Specifically we wish to examine the role that the event horizon plays and its connection to the global causal structure of spacetime. It is precisely the non-trivial causal structure related to event horizons that leads to problems with unitarity. Because we can "throw away" entangled particles in an absolute sense, we can have pure states evolving into mixed states. To this end we wish to argue that there are relatively few options for doing away with information loss other than doing away with event horizons. However, even if the spacetime does not contain event horizons it is possible to have a physically meaningful notion of a black hole. To replace event horizons, we propose a formalism for black holes that is already well established in the literature, based on marginally trapped surfaces.

To free us from this event horizon constraint we will use a quasi-local definition of the black hole, in terms of trapping horizons. At least at the semi-classical level, trapping horizons will be useful whatever the ultimate resolution of the issue of black hole evolution is.

To clarify what one might expect from black hole evaporation and give as clear a picture as possible, without solving the full back-reacting semi-classical Einstein equations with renormalised stress-energy tensor, we present here a simple, spherically symmetric picture making use of the Misner-Sharp mass function. In spherical symmetry the Misner-Sharp mass function can be interpreted as measuring the amount of energy within a sphere of areal radius $r$ at a time $\tau$. Therefore the Misner-Sharp mass function gives a reasonable, quasi-local definition to the concept of the curvature-producing energy contained within a black hole. We can use the Misner-Sharp mass to locate black hole horizons. It can give us insight into black hole thermodynamics and clarify our picture of what is happening when the black hole evaporates.

This paper is organised as follows. In section \ref{sec:infoloss} we review the standard arguments for how black holes can be used to turn pure states into mixed states. We emphasize the role played by the event horizon and by extension the non-trivial causal structure related to it, usually in terms of a spacelike singularity. In section \ref{trapping} we give a brief review of trapping horizons in spherical symmetry and how they give rise to black hole thermodynamics and Hawking radiation without any event horizon necessary. In section \ref{misnersharp} we describe how the Misner-Sharp mass can be used in spherical symmetry to follow the evolution of the black hole spacetime and how the various possible outcomes can be explained in terms of it.

\section{Information loss in black holes}
\label{sec:infoloss}

Information loss in black holes is related to the possibility of pure states evolving into fundamentally mixed states (for example see \cite{Hawking:1976,Mathur:2008wi}). A fundamentally mixed state is one that cannot be turned into a pure state by finer-graining or further measurements. In a black hole context, the mutation of pure states into mixed states comes about most strikingly in the context of Hawking radiation. Essentially virtual pair-production can lead to Hawking radiation if one of the virtual particles falls into the black hole, allowing the other particle to escape to infinity with real positive energy. The two particles are initially entangled with one another and are described by an entangled state
\beq \label{hawkingstate} |\psi\rangle = C e^{\gamma b_{1}^{\dagger}c_{1}^{\dagger}} | 0 \rangle, \eeq
where $C$ is a normalisation factor and $b_{1}^{\dagger}$ is a creation operator for the particles that escape to infinity and $c_{1}^{\dagger}$ is a creation operator for particles that fall into the black hole. For simplicity we can write a similar state as
\beq |\psi\rangle \ \frac{1}{\sqrt{2}}\left(|0\rangle_{b_{1}}\otimes |0\rangle_{c_{1}} + |1\rangle_{b_{1}}\otimes |1\rangle_{c_{1}}\right). \eeq
This state is similar to (\ref{hawkingstate}) if we just ignore all two and more particle states in the expansion of $e^{\gamma b_{1}^{\dagger}c_{1}^{\dagger}}$. This state is entangled because we cannot write it as a factored state of the form $|\psi\rangle_{1}\otimes |\psi\rangle_{2}$. While we still have both particles we can write the density operator for the system as a pure state
\bea \rho_{pure} = |\psi\rangle\langle \psi | & = & \hspace{0.25cm} \frac{1}{2}| 0\rangle_{b_{1}}\otimes | 0\rangle_{c_{1}} \langle 0 |_{b_{1}} \otimes \langle 0 |_{c_{1}} \nonumber \\ & & +\frac{1}{2}| 0\rangle_{b_{1}}\otimes | 0\rangle_{c_{1}} \langle 1 |_{b_{1}} \otimes \langle 1 |_{c_{1}} \nonumber \\ & & +\frac{1}{2}| 1\rangle_{b_{1}}\otimes | 1\rangle_{c_{1}} \langle 0 |_{b_{1}} \otimes \langle 0 |_{c_{1}} \nonumber \\ & & +\frac{1}{2}| 1\rangle_{b_{1}}\otimes | 1\rangle_{c_{1}} \langle 1 |_{b_{1}} \otimes \langle 1 |_{c_{1}}. \eea
However, if we `ignore' the particle that falls into the black hole, then we must sum or partial trace over the unknown state and we obtain a mixed state
\beq \rho_{mixed} = \frac{1}{2}| 0 \rangle_{b_{1}}\langle 0 |_{b_{1}} + \frac{1}{2}| 1 \rangle_{b_{1}}\langle 1 |_{b_{1}}. \eeq
This state is mixed because it cannot be written as single state $|\phi\rangle$ in the form $|\phi\rangle\langle\phi |$. In ordinary quantum systems this procedure of taking the partial trace over unmeasured subsystems is common and leads to `statistically' mixed states. The subsystem we focus on is described by a mixed state but this is only because we choose to focus on a part of the entire system. This subsystem is entangled with some other subsystem and the entire state, including the subsystem we choose not to measure, is still a pure state. If we chose to measure the full system we would recover the full pure quantum state.

However, when one of the subsystems has fallen over an event horizon or perhaps more importantly, has reached a boundary of spacetime and terminated, then there is no way, even in principle, that we could measure the state of the entire system, because one of the subsystems will never return to our past light cone. For all intents and purposes it is gone and we cannot choose to measure it sometime in the future. Therefore we must take a trace over the lost subsystem and the remaining state that is accessible to us will be a fundamentally mixed state.

It is this tracing over the unknowable degrees of freedom that leads to pure states turning into fundamentally mixed states. While both particles still exist one can claim that the full quantum state is pure. One could even claim that a particle that had fallen over the event horizon was still in principle measurable. However, if there is a spacetime boundary inside the black hole, such as a singularity, then the particle will reach the end of its worldline in finite proper time. After this it will cease to exist. If the black hole evaporates completely then the outgoing Hawking radiation particles will be entangled with nothing and they must represent a mixed state.

Whether one wants to associate the tracing over the unknowable states with them falling over the event horizon or with them falling into a singular spacetime boundary and ending is largely a matter of taste. Particles that fall over the event horizon cannot, by definition, return to the exterior asymptotic region, unless they move acausally. In the classical picture of a Schwarzschild black hole, particles that fall into black holes reach the singularity in a finite proper time and at the singularity they meet a spacetime boundary where their worldlines cannot be continued. In a more realistic, perhaps quantum description, whether their worldines actually come to an end at some finite parameter time depends on the full causal structure and the details of what one believes happens in the singular regions of general relativity.

A similar thing can happen if we allow naked timelike singularities. One of the entangled particles can fall into the singularity and disappear while the other particle escapes to infinity and becomes a fundamentally mixed state. This would appear to provide a mechanism for turning pure states into mixed states without event horizons, although the existence of a naked singularity would cause other problems as well. In some sense black holes are most commonly associated with information loss, only because the horizon readily creates the entangled particles and the spacetime singularity/boundary allows one of the particles to be erased.

The possibility of pure states turning into mixed states is therefore related to the possibility of the worldlines of particles coming to an end. This suggests that in order to understand whether information can be lost in black holes one must look at the possible endings of particle worldlines. There are in fact several possibilities for worldlines. Worldlines will either end at finite parameter time, in which case we would say they have reached a spacetime boundary, or they can be continued to infinite parameter time, in which case we would say that they end at infinity. If continued to infinite parameter time they can either end up in the same infinity of external observers or some other infinity, as occurs for example, for certain worldlines in the Reissner-Nordstr\"{o}m spacetime.

The are other possibilities for worldlines that are slightly less clearcut. One can consider the possibility that worldlines become undefined and it is meaningless to talk about their continuation. This may occur if spacetime because somehow `fuzzy' because of quantum effects. One can also consider the possibility that some new physics conspires to induce a breakdown of locality, or allows particles and observers to move along spacelike curves, so that causal Penrose diagrams are largely meaningless. This might occur for example if the interior boundary is defended by some membrane-like structure that prevents all freely-falling particles from passing through itself, along their geometric geodesic paths \cite{Yeom:2008qw}.

In most situations a spacelike boundary will give rise to an event horizon. This is the case with the spacelike singularity in the Schwarzschild solution. The spacelike boundary casts a shadow on light rays continued back from future infinity. Of course, spacelike singularities are not a necessary condition for event horizons. The Reissner-Nordstr\"{o}m spacetime has no spacelike boundaries but does have event horizons. The singularities in the Reissner-Nordstr\"{o}m spacetime are timelike and occur because of the disjoint null infinity structure.

If we want to avoid pure states evolving into mixed states we must not throw away the ingoing Hawking particles. If the ingoing particles reach a boundary in finite parameter time there will be hypersurfaces in the spacetime that their worldlines do not intersect and it is hard to argue that they should not be `thrown away'. There are several possibilities to avoid this outcome.

The first, perhaps most severe approach, is to deny the physical existence of black holes. This can be done, for example, by arguing that some non-trivial physics comes into play when matter attempts to collapse to form a black hole, as in for example \cite{Chapline:2000en,Mazur:2004fk,Vachaspati:2006ki,Mathur:2005zp}.

The second approach is to deny the existence of a spacetime boundary within the black hole that entangled particles can fall over. This is for example the purely classical proposal of Dymnikova \cite{Dymnikova:1997vk} and Hayward \cite{Hayward:2005gi} and the quantum geometrical proposal of \cite{Ashtekar:2005cj}. The possibility that one could have horizons without singularities seems to run foul of the singularity theorems but these can either be circumvented with non-global hyperbolicity \cite{Ansoldi:2008jw} or energy condition violation \cite{Hayward:2005gi} or a breakdown of the differentiable manifold picture \cite{Ashtekar:2005cj}. Violation of the energy conditions is perhaps not as drastic as it may sound, because Hawking radiation is known to violate nearly all of the energy conditions \cite{Visser:1996iw}.

\section{Trapping horizons}
\label{trapping}

If one follows the approach of denying the existence of spacetime boundaries within black holes, but one does not want to deny the existence of a black hole, one is left with the question of what one means by the black hole region. Traditionally, one defines the black hole using an event horizon. If there is no internal boundary within the black hole and no wormholes that connect the interior region to a different disjoint asymptotic region, since the worldlines of particles have to end somewhere, the only place left for them to end is at asymptotic future infinity. This means that technically there will be no event horizon.

If there is no event horizon, then how should one define the black hole? Here we suggest that one simply adopt the proposal of using some quasi-local definition of the black hole in terms of for example, a trapping horizon. Trapping horizons are defined quasi-locally in terms of the geometrical structure and are independent of the global causal structure. To see how trapping horizons provide an acceptable definition of black holes and even give rise to black hole thermodynamics and Hawking radiation, we give here a brief summary of trapping horizons in spherical symmetry. Further details can be found in the literature, see \cite{Nielsen:2005af} and references therein.

While spherical symmetry is not a reasonable assumption for astrophysical black holes it simplifies the discussion sufficiently to justify its adoption for `conceptual clarity'. Spherical symmetry is a reasonable assumption to study the formation and evaporation of black holes at a toy model level and it makes the analysis much simpler and clearer. Furthermore, in spherical symmetry, we can use the Misner-Sharp mass as a definition of quasi-local mass, which as we will see, also provides conceptual clarity to the issue of black hole evaporation.

A trapping horizon, more properly a future outer trapping horizon, is defined by Hayward \cite{Hayward:1993wb} as the closure of a three-surface which is foliated by marginal surfaces, for which $\theta_{l}=0$, and which, in addition, satisfies
\begin{enumerate}
\item[\it{i}.] $\theta_{n}<0$ (to distinguish between white holes
and black holes).
\item[\it{ii}.] $n^{a}\nabla_{a}\theta_{l}<0$ (to distinguish
between inner and outer horizons of, for example, the non-extremal Reissner-Nordstr\"{o}m solution).
\end{enumerate}\bigskip

Any spherically symmetric metric in four dimensions can be put in the form
\bea \d s^2 & = & - e^{-2\tilde{\Phi}(t,r)}\left(1-\frac{2m(t,r)}{r}\right)\d t^{2} + \nonumber \\ & & \frac{\d r^{2}}{\left(1-\frac{2m(t,r)}{r}\right)}+r^{2}\d\Omega^{2}, \eea
in so-called Schwarzschild coordinates, where $m(t,r)$ is immediately recognizable as the Misner-Sharp mass function. The Misner-Sharp mass function can be interpreted as the quasi-local mass contained within a sphere of radius $r$ at a time $t$. Unfortunately, this coordinate system is undefined at $r=2m(t,r)$ and cannot be continued through this hypersurface and cannot be relied upon for calculating properties at $r=2m$. A more appropriate coordinate system is perhaps the \Painleve-Gullstrand coordinate system

\bea \d s^{2} & = & -e^{-2\Phi(\tau,r)}\left(1-\frac{2m(\tau,r)}{r}\right)\d \tau^{2}+ \nonumber \\ & & 2e^{-\Phi(\tau,r)}\sqrt{\frac{2m(\tau,r)}{r}}\d\tau\d r + \d r^{2} + r^{2}\d\Omega^{2}. \eea
This coordinate system is regular at future horizons. The radial null geodesics for this metric can be easily found by setting $\d s = \d\Omega = 0$. For this we find
\beq \label{dtdtau} \frac{\d r}{\d\tau} = -e^{-\Phi(\tau,r)}\left( 1\pm \sqrt{\frac{2m(\tau,r)}{r}}\right), \eeq
where the plus sign denotes the ingoing geodesics. Thus we can find outgoing geodesics $l^{a}$ and ingoing geodesics $n^{a}$ with components
\beq \label{l} l^{a} = \left(e^{\Phi(\tau,r)},1-\sqrt{\frac{2m(\tau,r)}{r}},0,0\right), \eeq
\beq \label{n} n^{a} = \frac{1}{2}\left(e^{\Phi(\tau,r)},-1-\sqrt{\frac{2m(\tau,r)}{r}},0,0\right). \eeq
The factor of two ensures that the cross normalisation is the conventional $n^{a}l_{a} = -1$. Then we can compute
\beq \theta_{l} = \frac{2}{r}\left(1-\sqrt{\frac{2m(\tau,r)}{r}}\right), \eeq
\beq \theta_{n} = -\frac{1}{r}\left(1+\sqrt{\frac{2m(\tau,r)}{r}}\right). \eeq
We see that the expansion of $n^{a}$ is always negative and that at $r=2m(\tau,r)$ the expansion of $l^{a}$ is zero. We can also compute the value of $n^{a}\nabla_{a}\theta_{l}$ at $r=2m$
\beq \left(n^{a}\nabla_{a}\theta_{l}\right)_{H} = - \frac{(1-2m'_{H})}{r_{H}^{2}}\left(1+\frac{\dot{r}_{H}}{2e^{-\Phi_{H}}}\right), \eeq
where a subscript $H$ denotes functions to be evaluated at the horizon and we have used a dash to denote partial derivative with respect to $r$ and a dot to denote the partial derivative with respect to the time $\tau$ (here, since $r_{H}$ is only a function of $\tau$ it is actually an ordinary derivative).

For the horizon to be an outer horizon in a spacetime with a regular asymptotic region, we require $2m'_{H} < 1$, since $m(\tau,r)$ must be less than $r$ for large $r$. In addition, we can see from (\ref{dtdtau}) for the ingoing null geodesic $n^{a}$ that $\dot{r} = -2e^{-\Phi_{H}}$. Thus we see that we have a trapping horizon at $r=2m$ if the horizon is outer and not moving inwards faster than ingoing null geodesics.

The normal $N^{a}$ to the surface $r=2m$ has norm
\beq N^{a}N_{a} = -4\dot{m}_{H}^{2}e^{2\Phi_{H}}-4\dot{m}_{H}e^{\Phi_{H}}(1-2m'_{H}). \eeq
If $\dot{m}_{H}=0$ the trapping horizon will be a null hypersurface, and, assuming $1-2m'>0$, it will be a spacelike hypersurface if $\dot{m}>0$. For $-(1-2m')e^{\Phi} < \dot{m}<0$ the trapping horizon will be a timelike hypersurface. This opens the possibility that one can move along a causal curve from inside an evaporating horizon to the outside. For $\dot{m}<-(1-2m')e^{\Phi}$ the horizon is spacelike, but evaporating `faster than the speed of light' and so all timelike curves from a region just inside the horizon must move to the outside \cite{Nielsen:2005af}.

The surface $r=2m(r,t)$ does not however, define the location of the event horizon in a dynamical spacetime. The event horizon is always a null surface and so the spherically symmetric trapping horizon at $r=2m$ can only be an event horizon if $\dot{m}=0$ (note however that this is necessary but not sufficient). To find the event horizon, firstly one would need an explicit form for $m(r,t)$ and then one would look for radial null vectors that are not able to reach infinity by propagating them outwards from the centre of the spacetime. 

It has been well-known for a long time that locally defined horizons can give rise to equations of black hole thermodynamics \cite{Collins:1992}. At the horizon we have
\beq \label{firstlaw} \frac{\partial m}{\partial \tau} = \frac{1}{8\pi}\frac{(1-2m')}{2r}\frac{\d A}{\d \tau}, \eeq
where $m'=\frac{\partial m}{\partial r}$ and $\tau$ is a parameter labeling `time-slicings' of the horizon. This equation has the same form as the first law of black hole thermodynamics $\delta m = \frac{1}{8\pi}\kappa\;\delta A$ with a surface gravity that agrees with other definitions of surface gravity \cite{Nielsen:2007ac}. In order to obtain a version of the second law we can just compute $G_{ab}l^{a}l^{b}$, where $G_{ab}$ is the Einstein tensor. This gives
\beq G_{ab}l^{a}l^{b} = \frac{2e^{\Phi}}{r^{2}}\frac{\partial m}{\partial \tau}\sqrt{\frac{2m}{r}} - \frac{2}{r}\frac{\partial\Phi}{\partial r}\left(1-\sqrt{\frac{2m}{r}}\right)^{2}. \eeq
Rearranging, and imposing (\ref{firstlaw}) at $r=2m(\tau,r)$ we find
\beq \label{secondlaw} \frac{\partial A}{\partial \tau} = \frac{16\pi r^{3}e^{-\Phi}}{1-2m'} G_{ab}l^{a}l^{b}. \eeq
Thus we see that the area of the horizon $A$ is increasing if $G_{ab}l^{a}l^{b} > 0$. By the Einstein equations we can write this condition as $T_{ab}l^{a}l^{b} > 0$, which is just the null energy condition (NEC). The area of the horizon is increasing if the NEC is satisfied and can decrease only if the NEC is violated.

Furthermore, using the Parikh-Wilczek tunneling approach to Hawking radiation we may expect to have Hawking radiation produced in the vicinity of the trapping horizon \cite{Visser:2001kq,Di Criscienzo:2007fm,Clifton:2008sb}. This is perhaps not surprising, since the effect is just a result of quantum field theory on a curved background and quantum field theory is a local theory that is unlikely to depend on global properties like an event horizon \cite{Nielsen:2008dj}.

Consider the equation for a massless scalar field on a curved background
\beq \frac{\hbar^{2}}{\sqrt{-g}}\partial_{a}\left( g^{ab}\sqrt{-g}\partial_{b}\right)\phi = 0. \eeq
We look for solutions of the form $\phi = exp(-iS(\tau, r)/\hbar)$. Taking this limit as $\hbar \rightarrow 0$, to lowest order this equation gives the Hamilton-Jacobi equation
\beq g^{ab}\partial_{a}S\partial_{b}S = 0. \eeq
Invoking the geometrical optics approximation
\beq S(\tau,r) = \omega t - \int k(r)\d r, \eeq
this equation gives
\beq \omega^{2} + 2e^{-\Phi}\sqrt{\frac{2m}{r}}\omega k - e^{-2\Phi}\left(1-\frac{2m}{r}\right)k^{2} = 0. \eeq
Rearranging gives
\beq k = \pm\frac{\omega e^{\Phi}}{1\mp\sqrt{\frac{2m}{r}}}. \eeq
The plus sign denotes the outgoing modes and the negative sign denotes the ingoing modes.
The outgoing modes contain a pole at the horizon $r=2m$.
\beq S = \omega t + \frac{2r_{H}\omega e^{\Phi_{H}}}{\left(1-2m'_{H}\right)}\int\frac{\d r}{\left(r-r_{H}\right)}. \eeq
The integral can be performed by deforming the contour into the lower half of the complex plane, which gives a complex part to $S$
\beq \textrm{Im}S = \frac{4\pi r_{H}\omega e^{\Phi_{H}}}{\left(1-2m'_{H}\right)}. \eeq
At this level of approximation this corresponds to thermal radiation with a temperature
\beq T = \frac{1}{2\pi}\frac{e^{-\Phi_{H}}}{2r_{H}}\left(1-2m'_{H}\right), \eeq
which agrees with the calculations in \cite{Nielsen:2007ac}.

\section{Black hole evaporation with the Misner-Sharp mass}
\label{misnersharp}

\begin{figure}[ht]
\includegraphics[scale=0.5]{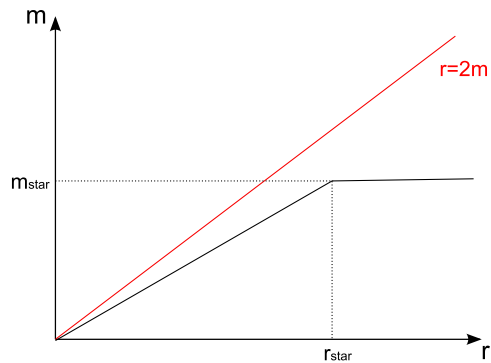}
\caption{The radial mass profile as it might appear for a star of radius $r_{star}$ and mass $M_{star}$.}
\label{mstar}
\end{figure}

\begin{figure}[hb]
\includegraphics[scale=0.5]{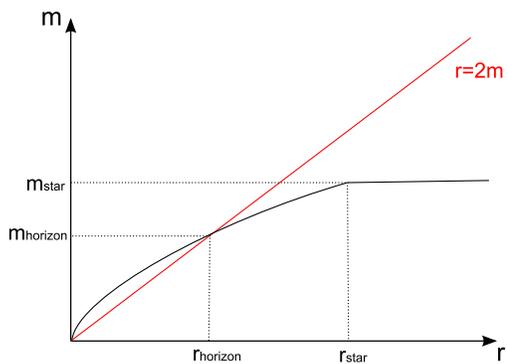}
\caption{As the star collapses, a horizon forms, initially with a small radius.}
\label{mstarII}
\end{figure}

\begin{figure}[ht]
\includegraphics[scale=0.5]{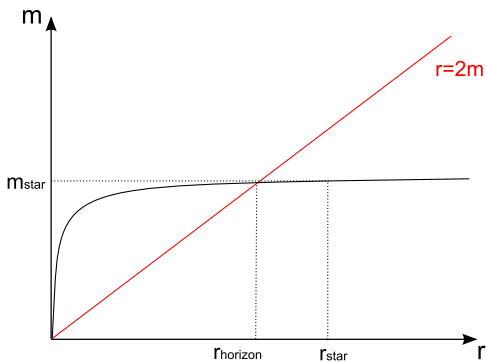}
\caption{As the collapse proceeds the area of the horizon grows.}
\label{mstarIII}
\end{figure}

\begin{figure}[ht]
\includegraphics[scale=0.5]{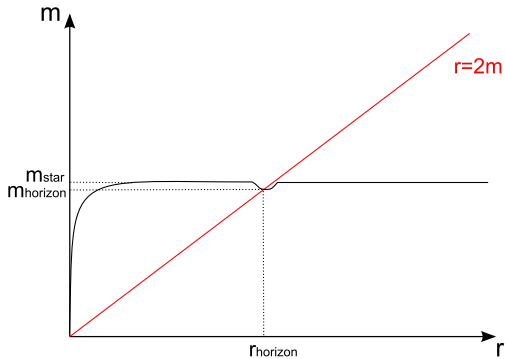}
\caption{Hawking radiation occurs in the vicinity of the horizon. This means that positive energy flows outwards and a corresponding negative energy flows inwards. The size of this effect for a macroscopically sized black hole is greatly exaggerated here.}
\label{mstarIV}
\end{figure}

\begin{figure}[ht]
\includegraphics[scale=0.5]{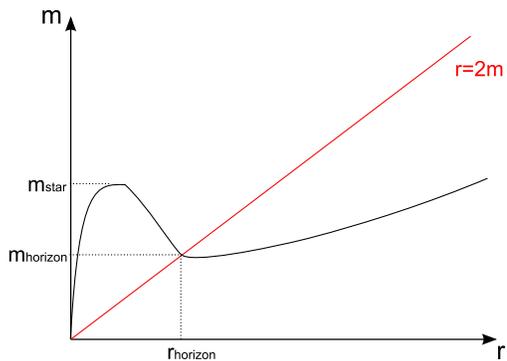}
\caption{As the Hawking evaporation proceeds, the horizon area shrinks. The flux of positive energy moves outwards to greater and greater radius, and the flux of negative energy moves inwards to the centre.}
\label{mstarV}
\end{figure}

\begin{figure}[ht]
\includegraphics[scale=0.5]{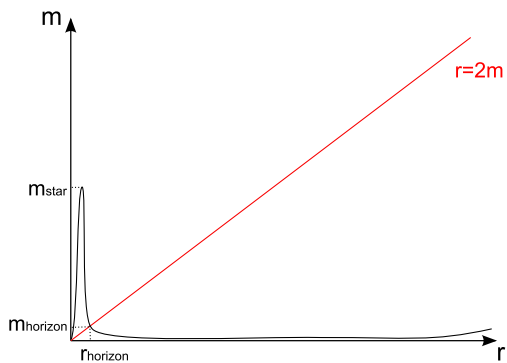}
\caption{Eventually, sufficient energy has evaporated away from the black hole to leave the intermediate spacetime effectively flat. An observer in this region would feel almost no gravitational field from the collapsed star. Whether the interior region is flat as well depends on the physics at small $r$}
\label{mstarVI}
\end{figure}

\begin{figure}[ht]
\includegraphics[scale=0.5]{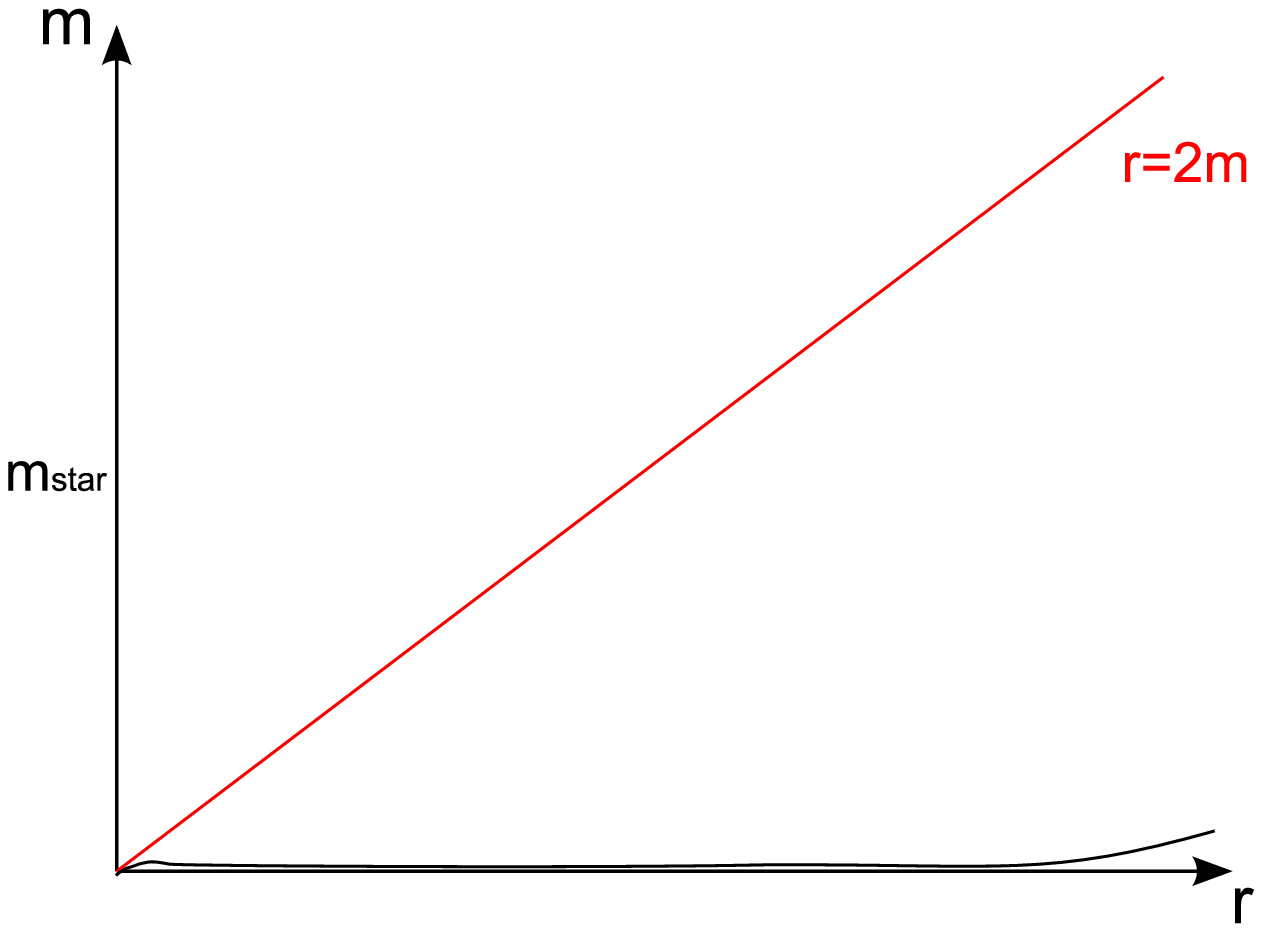}
\caption{How the ingoing flux of negative energy interacts with the interior mass of the star depends on the details of the calculation at small radii. This is the region where the matter of the originally collapsed star is compressed to great densities and the region where one might expect quantum gravitational effects to be important. If the negative energy and original star matter annihilate completely, one will end up with a spacetime that is effectively flat Minkowski space.}
\label{mstarVII}
\end{figure}

The Misner-Sharp mass \cite{Misner:1964je} can be very useful for investigating the behaviour of black holes \cite{Hayward:1994bu}. For a spherically symmetric metric, using the Einstein equations, we have
\beq \label{T00} T_{0}^{\hspace{0.1cm}0} = \frac{m'}{4\pi r^{2}}, \eeq
where $m'$ denotes partial differentiation with respect to the coordinate $r$. This will correspond to the local energy density $\rho$ as measured by an observer with constant coordinate $r$. Hypersurfaces of constant \Painleve-Gullstrand time $\tau$ are spacelike since the normal to the surface $\nabla_{a}\tau$ satisfies
\beq \nabla^{a}\tau\nabla_{a}\tau = -e^{2\Phi(\tau,r)}. \eeq
On a $\tau = \mathrm{constant}$ hypersurface $\sigma$ we can write a quasi local Misner-Sharp mass as
\beq M = \int_{\Sigma} \rho \sqrt{1-2m/r} \d V, \eeq
where $\d V$ is the volume element on the hypersurface. In the words of Misner and Sharp \cite{Misner:1964je}, the function $m$ includes contributions from the kinetic energy and gravitational potential energy. Since the volume element is just $r^{2}\sin\theta/\sqrt{1-2m/r}$ we obtain
\beq M = \int m' \d r = m. \eeq
The Misner-Sharp mass reduces to the Arnowitt-Deser-Misner (ADM) mass at spacelike infinity and reduces to the Bondi-Sachs mass at null infinity. In the Newtonian limit is also gives the Newtonian mass to first order \cite{Hayward:1994bu}. 

We can also see from (\ref{T00}) that if the local energy density is negative then $m'$ will also be negative. A negative energy density corresponds to a violation of the Weak Energy Condition (WEC). Using the Page approximation \cite{Page:1982fm} for the renormalised energy-momentum tensor in the Hartle-Hawking vacuum of a Schwarzschild black hole we know that Hawking radiation is expected to violate the WEC and NEC \cite{Visser:1996iw}. Thus we might expect, in a fully backreacting spacetime, that in the region near the black hole horizon $m'$ should be negative. In fact, we know that the null energy condition must be violated in the vicinity of an evaporating, area decreasing horizon (\ref{secondlaw}). This also means that the surface gravity and therefore temperature of the black hole can be greater than the corresponding temperature of a Schwarzschild black hole with the same mass \cite{Nielsen:2007ac}.

The Misner-Sharp mass allows us to construct the following heuristic picture of black hole formation and evaporation. By taking equal time ``snapshots" of the Misner-Sharp mass function on various constant $\tau$ coordinate slices we can imagine the evolution of a star as it undergoes gravitational collapse to form a black hole and how Hawking radiation evaporates away the mass-energy of the star.

An important feature of this picture is the $r=2m$ condition that allows us to see the location of the trapping horizon boundary of the black hole. This is indicated in the plots by a red line. Initially, Fig.\ref{mstar}, the star has some stellar mass profile, without necessarily any prejudice as to whether the density goes like $\rho ~ r^{-2}$ as depicted here, and vacuum outside. At this time, nowhere is the condition $r=2m$ reached. As the star collapses, a horizon forms, initially with a small radius, Fig.\ref{mstarII}. As the collapse proceeds the area of the horizon grows, Fig.\ref{mstarIII}. Hawking radiation is expected to occur in the vicinity of the horizon, Fig.\ref{mstarIV}. This means that positive energy flows outwards and a corresponding negative energy flows inwards. The size of this effect for a macroscopically sized black hole is greatly exaggerated here. As the Hawking evaporation proceeds, the horizon area shrinks, Fig.\ref{mstarV}. The flux of positive energy moves outwards to greater and greater radius, and the flux of negative energy moves inwards to the centre. Eventually, sufficient energy has evaporated away from the black hole to leave the intermediate spacetime effectively flat, Fig.\ref{mstarVI}. An observer in this region would feel almost no gravitational field from the collapsed star.

How the ingoing flux of negative energy interacts with the interior mass of the star depends on the details of the calculation at small radii. This is the region where the matter of the originally collapsed star is compressed to great densities and the region where one might expect quantum gravitational effects to be important. If the negative energy and original star matter annihilate completely, one will end up with a spacetime that is effectively flat Minkowskian space, Fig.\ref{mstarVII}.

If we want to view the collapse purely classically then the boundary term at $r=0$ is crucial. In the above plots we have imposed $m(\tau,r=0) =0$ as was done, for example, in \cite{Hayward:2005gi}. To investigate the behaviour at $r=0$ we can compute the Kretschmann scalar for the simple static case with $\Phi(\tau,r)=0$. If we take $m(r) \sim r^{n}$ to leading order we get
\beq R_{abcd}R^{abcd} \sim \left(n^{4}-6n^{3}+17n^{2}-20n + 12\right)\frac{r^{2n}}{r^{6}}. \eeq
This quartic equation in $n$ has no real roots and therefore if $m(r) \sim r^{n}$ with $n \leq 3$ we will get a singularity at $r=0$ in the Kretschmann scalar. Of course, a finite Kretschmann scalar is necessary but not sufficient for there to be no singularity at $r=0$.

\begin{figure}[ht]
\includegraphics[scale=0.5]{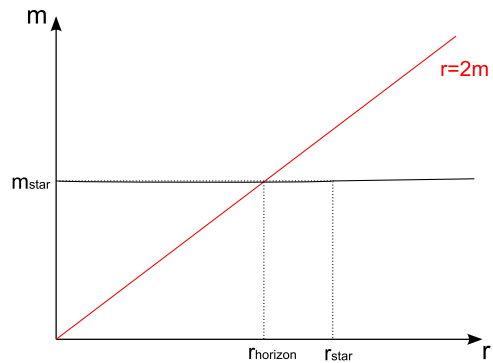}
\caption{For the Schwarzschild solution, the Misner-Sharp mass function is a constant, $M$, for all values of $\tau$ and $r$.}
\label{mstarschwarz}
\end{figure}

\begin{figure}[ht]
\includegraphics[scale=0.5]{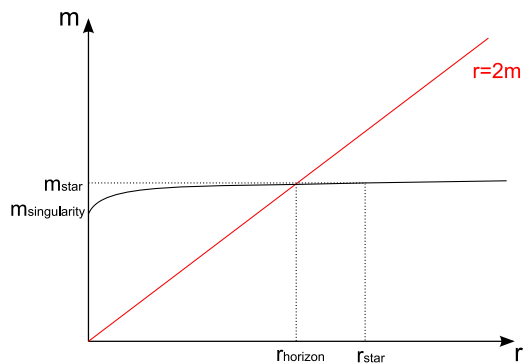}
\caption{We can imagine the possibility that there is a finite amount of mass contained within a sphere of radius zero. In this case there will be a singularity at $r=0$.}
\label{mstarsing}
\end{figure}

For the Schwarzschild solution we have the condition $m(\tau,r) = M$ for all $\tau$ and $r$, Fig.\ref{mstarschwarz}. This suggests that the collapse may proceed in the way depicted in Fig.\ref{mstarsing}. If $m>0$ at $r=0$ then we obtain a spacelike singularity at $r=0$. Indeed, it is even possible that we could have $m<0$ at $r=0$ in which case we would have an untrapped (weakly naked) timelike singularity \cite{Hayward:1994bu}. 

\begin{figure}[ht]
\includegraphics[scale=0.5]{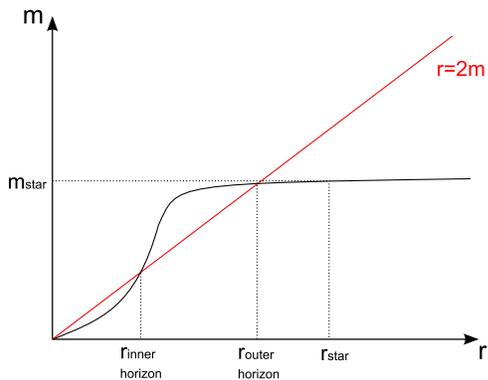}
\caption{An inner and outer horizon. The region between the inner and outer horizons contains trapped surfaces, while the region inside the inner horizon does not.}
\label{mstardouble}
\end{figure}

The final thing that it is easy to show using the Misner-Sharp mass is the creation of an inner and outer horizon, as for example considered in \cite{Hayward:2005gi}. The simple condition for this to occur is that
\beq m' > \frac{1}{2}, \eeq
at the inner horizon where $r=2m$. In Hayward's scenario the trapping horizon forms initially as an extremal horizon at a finite radius. Then the inner and outer horizons separate to form a trapped region (with the region close to $r=0$ untrapped) and then the inner and outer horizons join up again forming an instantaneously extremal horizon before disappearing altogether. This is shown in Fig.\ref{mstardouble}. The region between the inner and outer horizons contains trapped surfaces, while the region inside the inner horizon does not.
\section{Conclusion}

We have seen how the causal structure of spacetime plays an important role in the semi-classical information loss argument. In this way, information loss is very closely tied up with defining black holes via event horizons. One often reasons that in order to have a black hole one must have an event horizon. In order to have an event horizon one must have a non-trivial causal structure. If one has a non-trivial causal structure then one always has the possibility of information loss.

Current notions about what constitutes a black hole are, of course, a result of the history of the development of the subject. The early days of Finkelstein, Wheeler and others started out with a purely classical picture of a black hole based largely on exact solutions, which by their very nature, required a large amount of symmetry in order to be discovered as solutions of the Einstein equations. Penrose, Hawking and others then laid down the fundamental concepts of what it means to be a black hole as a region that one can never escape from and developed techniques for dealing with situations beyond exact solutions. Then the laws of black hole thermodynamics were discovered, still in the context of black holes that eventually settle down to just being black before finally, Hawking discovered black hole radiation and the possibility that black holes could disappear arose. Had Hawking radiation been discovered at an earlier stage, the community might not be so insistent on defining black holes in terms of event horizons.

We have presented a picture that hopefully clarifies the issue of how event horizons are related to information loss. It is not based on an exact solution to the semi-classical Einstein equations but gives the properties one might expect from a full solution. It is also important to recognise what assumptions have gone into the above picture.

Firstly, we have assumed spherical symmetry. This greatly simplifies the formalism and allows us to use the Misner-Sharp mass as a quasi-local measure of energy. It is of course, not a valid model for describing astrophysical black holes, but it is a valid assumption to study black hole evaporation in its purest form. It is not expected that deviations from spherical symmetry will qualitatively change the physics of black hole evaporation.

Secondly, we have chosen to define the black hole in terms of a trapping horizon. This is not the usual definition of a black hole, but arises from the desire to define black holes without appealing to event horizons. There are many issues that remain to be clarified about the use of trapping horizons and they may ultimately turn out not to be the optimal choice. Questions about the uniqueness of trapping horizons have not been studied much in the literature and one can certainly ask questions about how a spacelike trapping horizon should intersect a given ``constant time" spatial hypersurface.

Thirdly, we have of course assumed that it is meaningful to talk about the matter that falls into the black hole, and this irrespective of whether it is the original collapsing star or matter that is subsequently accreted. This aspect is often ignored in other studies, partly because whatever falls through the event horizon, can, by definition, be ignored in terms of its effect of the outside region and secondly because there is in some quarters an implicit belief that whatever the state of the interior black hole is, it can somehow be read off from physics at the horizon.

Here we argue instead that the collapse of the central matter is essential to determining the causal structure of the spacetime and this is ultimately responsible for determining whether one can turn pure states into mixed states. Black holes cannot be truly understood until we understand what happens to all the neutrons (and protons and other particles) of the original star as they collapse under gravity to ever higher densities.

It is perhaps tempting to think that since quantum gravity effects are only likely to come into play in regions of high curvature, that the existence of an event horizon, since it occurs in a region of relatively low curvature, is unlikely to be affected. However, it is important to remember that the event horizon is not defined in terms of the local geometry, but causally in terms of the global causal structure. If the collapse is somehow halted or a singularity avoided then small changes in the microscopic central region can have large effects on the global causal structure, in the same way that adding one electron's worth of charge to an otherwise supermassive black hole can change its causal structure from Schwarzschild to Reissner-Nordstr\"{o}m. This change, while insignificant in terms of local physics, corresponds to a radical change of the causal structure, turning spacelike singularities into timelike singularities amongst other things.

The picture presented above also downplays several features commonly associated with information loss in black holes. Firstly, the uniqueness theorems (often called no-hair theorems) that hold for stationary, asymptotically flat, electro-vac, Einstein-Maxwell black hole spacetimes \cite{Heusler:1998ua} play little role. Nor does the issue of whether the Hawking radiation is exactly thermal or not. All that is essential is that previously entangled particles become entangled with nothing when their entangled ``partners" are thrown away. One could imagine a black hole spacetime with infinitely many independent measurable charges at infinity \cite{Leith:2007bd}, if the spacetime still has a spacetime boundary inside the black hole, over which entangled particles can be lost, then there is still the possibility of information loss. One could also imagine a situation with a single pair of entangled particles. If one of them is lost to a spacetime boundary the other particle will be in a fundamentally mixed state, and a single particle can hardly be said to describe a thermal spectrum.

Another related feature to uniqueness is the possibility of retrodicting the original state that collapsed to form the black hole from measurements of the final state. This should also be conceptually separated from the loss of unitarity in black holes. Such loss of retrodictablility occurs in ordinary particle physics where one knows that the S-matrix is perfectly unitary. For example, since both electron-positron pairs and muon-anti muon pairs can co-annihilate to produce high energy photons, if one just observes the high energy photons at late time, one cannot retrodict whether they arose due to electron-positron co-annihilation or a different particle-antiparticle pair.

Another implicit assumption that one often meets is that the evaporating Hawking radiation somehow is linked to the originally collapsed matter. This is because energy is conserved and if the Hawking radiation is carrying away energy, then that energy must somehow be the energy of the original star. What we have shown using the Misner-Sharp mass function is that this need not be the case. One can think of the Hawking radiation as a pure quantum field effect on a curved background. It is the spacetime that it evaporating and not the collapsed star. The original mass of the star can remain clumped at the centre (in some as yet unknown state) while the Hawking effect proceeds to produce an outgoing flux of energy from the vicinity of the horizon.

To conclude we highlight some of the outstanding questions that remain to be answered. What happens to the collapsing matter as it gets compressed to ever higher densities? What is the physics that is truly responsible for generating the Hawking radiation and does it effect the evolution of the black hole, and how does the negative energy produced in the Hawking radiation interact with the collapsed, highly compact remnant of the star? We believe that the answers to these questions will be needed before we can truly say we understand black hole evolution.

\end{document}